\documentclass[10pt,twocolumn,letterpaper]{article}

\usepackage[pagenumbers]{cvpr} 

\usepackage{graphicx}
\usepackage{amsmath}
\usepackage{amssymb}
\usepackage{booktabs}
\usepackage{float}

\usepackage[inline]{enumitem}
\usepackage{multirow}
\usepackage{gensymb}
\usepackage{siunitx} 
\ifdefined\qty\else
  \ifdefined\NewCommandCopy
    \NewCommandCopy\qty\SI
  \else
    \NewDocumentCommand\qty{O{}mm}{\SI[#1]{#2}{#3}}
  \fi
\fi
\ifdefined\unit\else
  \ifdefined\NewCommandCopy
    \NewCommandCopy\unit\si
  \else
    \NewDocumentCommand\unit{O{}m}{\si[#1]{#2}}
  \fi
\fi

\newcommand{\aid}{AiD Regen\xspace}
\newcommand{\deeplab}{DeepLabv3\xspace}
\newcommand{\rokitdata}{RKT-DFU-800\xspace}

\usepackage{hyperref}
\hypersetup{breaklinks,colorlinks}

\usepackage[capitalize]{cleveref}
\crefname{section}{Sec.}{Secs.}
\Crefname{section}{Section}{Sections}
\Crefname{table}{Table}{Tables}
\crefname{table}{Tab.}{Tabs.}

\def\cvprPaperID{9662} 
\def\confName{CVPR}
\def\confYear{2022}

\begin{document}

\title{Generating 3D Bio-Printable Patches Using Wound Segmentation and Reconstruction to Treat Diabetic Foot Ulcers}

\author{Han Joo Chae \quad Seunghwan Lee \quad Hyewon Son \quad Seungyeob Han \quad Taebin Lim\\
ROKIT Healthcare, Inc.\\
{\tt\small \{hanjoo.chae, seunghwan.lee, hyewon.son, seungyeob.han, taebin.lim\}@rokit.co.kr}
}


\twocolumn[{%
\renewcommand\twocolumn[1][]{#1}%
\maketitle
\begin{center}
    \centering
    \captionsetup{type=figure}
    \begin{subfigure}{0.23\linewidth}
        \includegraphics[width=\textwidth]{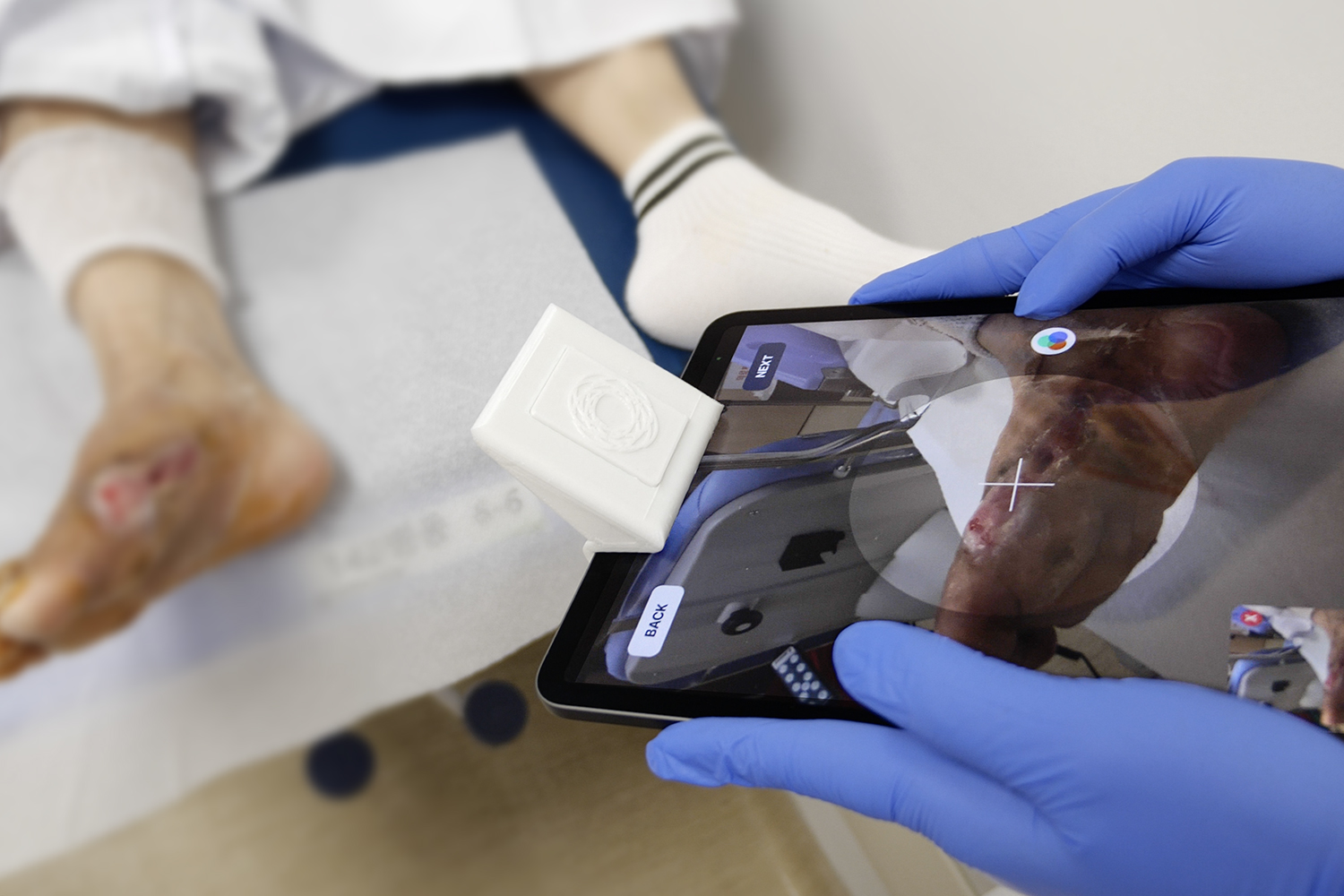}
        \caption{Wound capturing}
        \label{fig:teaser-a}
    \end{subfigure}
    \hspace{0.1pt}
    \begin{subfigure}{0.23\linewidth}
        \includegraphics[width=\textwidth]{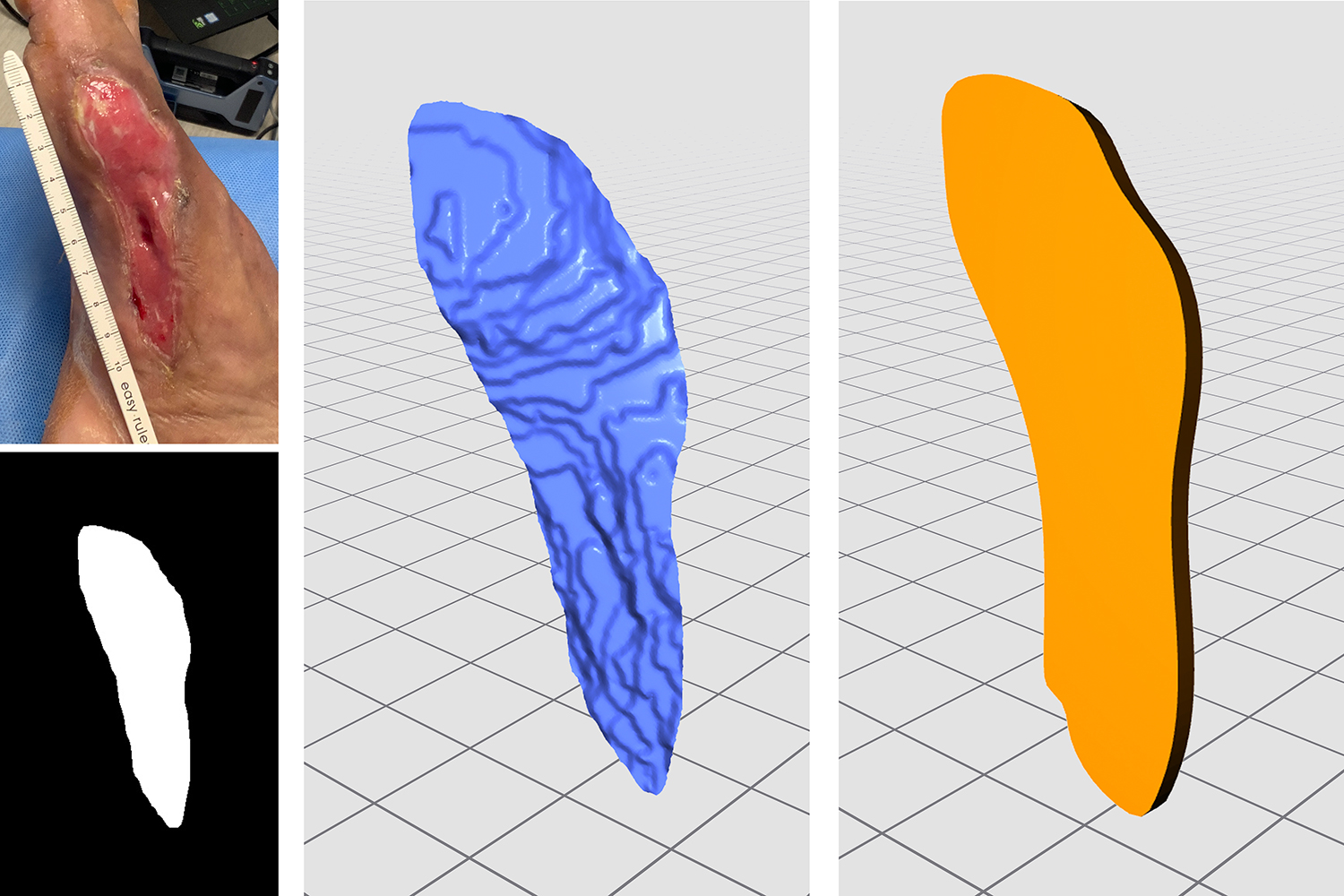}
        \caption{3D patch generation}
        \label{fig:teaser-b}
    \end{subfigure}
    \hspace{0.1pt}
    \begin{subfigure}{0.23\linewidth}
        \includegraphics[width=\textwidth]{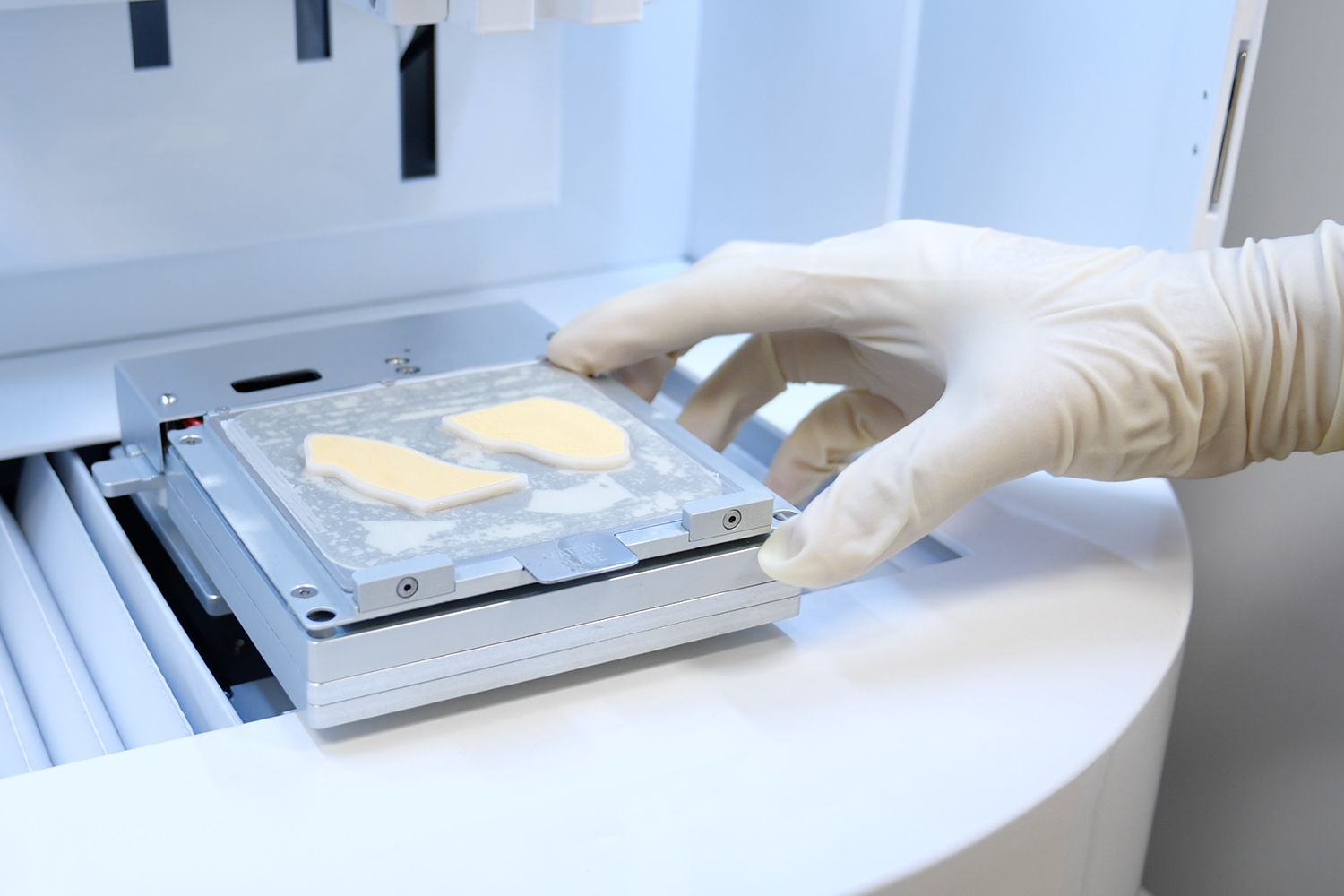}
        \caption{3D bio-printing}
        \label{fig:teaser-c}
    \end{subfigure}
    \hspace{0.1pt}
    \begin{subfigure}{0.23\linewidth}
        \includegraphics[width=\textwidth]{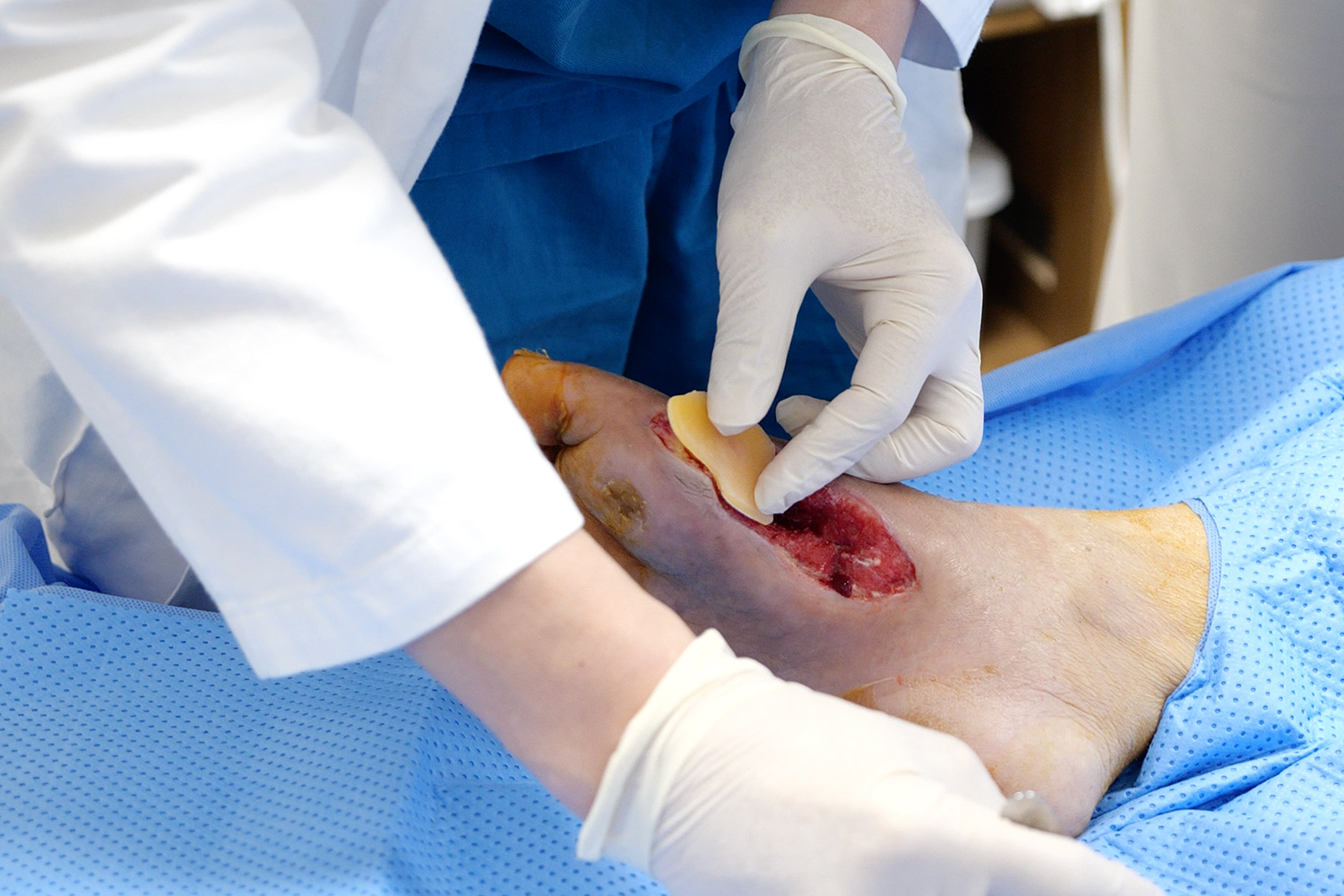}
        \caption{Applying the patch}
        \label{fig:teaser-d}
    \end{subfigure}
    \caption{The procedure of the DFU treatment includes (a) wound image capturing, (b) hybrid 3D patch generation combining 2D semantic segmentation and 3D reconstruction, (c) printing the patch using a 3D bio-printer, and (d) applying the printed patch to the wound.}
    \label{fig:teaser}
    
\end{center}%
}]


\begin{abstract}
   We introduce AiD Regen, a novel system that generates 3D wound models combining 2D semantic segmentation with 3D reconstruction so that they can be printed via 3D bio-printers during the surgery to treat diabetic foot ulcers (DFUs). AiD Regen seamlessly binds the full pipeline, which includes RGB-D image capturing, semantic segmentation, boundary-guided point-cloud processing, 3D model reconstruction, and 3D printable G-code generation, into a single system that can be used out of the box. We developed a multi-stage data preprocessing method to handle small and unbalanced DFU image datasets. AiD Regen's human-in-the-loop machine learning interface enables clinicians to not only create 3D regenerative patches with just a few touch interactions but also customize and confirm wound boundaries. As evidenced by our experiments, our model outperforms prior wound segmentation models and our reconstruction algorithm is capable of generating 3D wound models with compelling accuracy. We further conducted a case study on a real DFU patient and demonstrated the effectiveness of AiD Regen in treating DFU wounds.
\end{abstract}


\section{Introduction}
\label{sec:intro}

Diabetic foot ulcer (DFU) is the most costly and devastating complication of diabetes mellitus, which affects \qty{15}{\%} of diabetic patients during their lifetimes~\cite{dfu-management-review}. Poor blood circulation and high glucose levels decrease diabetic patients' ability to heal, and every $20$ seconds, a leg is lost through amputation due to infection and necrosis~\cite{dfu-systematic-review}. With the urgent need for a novel treatment method, the integration of 3D printing and bio-materials is paving the way toward devising innovative solutions in regenerative medicine~\cite{3D-bioprinting-frontier}. Kesavan et al.~\cite{dfu-india} have applied 3D bio-printed patches, that consist of minimally manipulated extracellular matrix (MA-ECM) harvested from autologous fat tissues, to DFU wounds and shown a complete healing with re-epithelialization only within about four weeks.

However, such use of 3D bio-printing technology is not accessible for many medical practitioners as it requires a complex process of 3D scanning and modeling. Conventional medical scanning methods, such as CT and MRI scans, often demand large, special devices and environments as well as experts who can operate them. In addition, the results of these scans usually include unnecessary parts and peripherals so that hospitals need to hire CAD engineers for post-processing the results or rely on outsourcing; either way, cost increases are inevitable. Also, since treatments of skin wounds like DFU are usually followed by debridement, a surgical removal of damaged tissue, the 3D patch creation must be done as quickly as possible to prevent debrided wounds from contamination. External post-processing and outsourcing might not be the best solutions in such time sensitive situations. Though portable industrial 3D scanners~\cite{einscan-pro-2x} lowered spatial and technical barriers, they still require users to understand correct scanning methods, and the post-processing needs remain the same.

3D object reconstruction using RGB-D images~\cite{3d-recon-rgbd} has been actively studied to automate modeling tasks; segmenting each part of an object can be done in real-time~\cite{FG-net}. Unfortunately, however, most existing wound datasets consist only of 2D images, and available medical data are often small or fragmented due to their sensitive nature; that is, preparing a large RGB-D or 3D wound training dataset might be difficult. Hybrid approaches, that combine features extracted from 2D images with depth data, could be good solutions to this as they have shown convincing performance in reconstructing wounds~\cite{wound-measurement, laser-wound-recon, wound-measurement-multimodal}.

In this work, we introduce \emph{\aid}, a novel system that generates 3D models based on the shape and geometry of wounds combining 2D image segmentation and 3D reconstruction so that they can be printed via 3D bio-printers during the surgery to treat DFUs. \aid seamlessly binds the full pipeline, which includes RGB-D image capturing, semantic segmentation, boundary-guided point-cloud processing (BGPCP), 3D model reconstruction, and 3D bio-printable G-code generation, into a single system that can be used out of the box. We also develop a multi-stage data pipeline to handle small and unbalanced characteristics of a DFU image dataset. Reflecting requirements of medical practitioners, \aid's human-in-the-loop (HITL) machine learning (ML) interface enables clinicians to not only create 3D regenerative patches with just a few number of touch interactions, but also customize and confirm wound boundaries. We evaluate the performance of our 2D segmentation and 3D model generation methods through a series of experiments and demonstrate the effectiveness of \aid by conducting a case study on a real DFU patient.
\section{Related Work}
\label{sec:relatedwork}

\subsection{3D Reconstruction and Segmentation}

Along with the advancements in RGB-D cameras, reconstructing real-world environments to 3D models has been actively studied~\cite{3d-recon-rgbd}. In addition, efforts have been exerted to separate each object in the given scene as in RGB-D based semantic segmentation~\cite{multi-task-learning, depth-encoding, multi-scale, novel-network, 3d-fusion, post-processing}. However, these research using public 3D semantic segmentation dataset are yet immature and their performances are reported at below \qty{50}{\%} mIoU~\cite{dl-3d-part-seg-survey}. Moreover, gathering 3D segmentation data is difficult since it requires high quality sensors as well as 3D domain expertise.

In terms of medical applications, hybrid approaches, that combine 2D image features with depth data to reconstruct 3D objects, have been most commonly used. Filko et al.~\cite{wound-measurement} reconstructed a 3D scene using the marching cubes algorithm and segmented wounds using color histograms and k-Nearest-Neighbor algorithm. Gutierrez et al. ~\cite{wound-measurement-multimodal} used deep learning methods to perform semantic segmentation on DFU wounds and merged the results with the 3D foot model reconstructed from a sequence of RGB images. Our \aid system also takes a hybrid approach to exploit well-established, deep-learning-based 2D semantic segmentation to find accurate boundaries of DFU wounds and use them to create 3D regenerative patches.

\subsection{2D Semantic Segmentation}
Since fully convolutional network has been proposed~\cite{fcn}, deep learning approaches based on convolutional neural network (CNN) have shown great successes in semantic segmentation. Spatial pyramid pooling~\cite{sppnet, pspnet} has shown its effectiveness in obtaining details of objects, and DeepLab~\cite{deeplabv3+} has been one of the most successful networks utilizing this technique. DeepLab applies atrous convolution to get multi-scale features and then concatenate the resulting features; this helps with getting feature maps from various receptive fields while keeping the high computational efficiency.

Attention model~\cite{attention-seg-original, senet-channel-attention, deeplabv3+-attention} also has been successful in semantic segmentation as it weighs the importance of scale or channel. SENet~\cite{senet-channel-attention} utilized channel attention to adaptively recalibrate channel-wise feature responses based on inter-dependencies between channels. Chen et al.~\cite{attention-seg-original} utilized scale attention to focus on weighting multi-scale features.

Transformer models are also actively being applied to semantic segmentation tasks~\cite{transformer-seg1, transformer-seg2}. These research use self-attention mechanisms rather than CNNs as their main structures. While the Transformer models are competent in semantic segmentation, they perform worse than CNN-based networks when the training dataset is not large enough~\cite{transformer-data-inefficiency}.

In contrast to general objects, relatively little attention has been paid towards wound segmentation. Wang et al.~\cite{wound-seg3} systematically compared VGG16, SegNet, U-Net, Mask-RCNN, and MobileNetV2 networks on wound segmentation tasks and reported that MobileNetV2 showed the best performance. While there is yet no large public dataset specifically designed for DFU segmentation, a few models have been proposed~\cite{dfu-seg1,dfu-seg2,dfu-seg3} and CNN-based models have shown the most reliable performances due to their data-efficient nature.

In this paper, we build a new DFU dataset (\rokitdata) consisting of real-world DFU wound images taken by clinicians during surgeries or examinations and overcome the nature of unbalanced real-world data with wound-specific augmentation techniques. Inspired by the work of Azad et al.~\cite{deeplabv3+-attention}, we apply the attention mechanism to \deeplab to preserve details of wound boundaries.

\subsection{Medical Machine Learning Systems}

While a proliferation of studies on individual ML tasks, such as semantic segmentation ~\cite{deeplabv3+, hrnet, point-transformer, sementicseg-deeplearning} and 3D reconstruction~\cite{img-based-3d-obj-recon, review-3d-recon-indoor}, has been published each year, relatively little studies have focused on applying and adapting those techniques to specific real-world problems, which require significant algorithmic and engineering work~\cite{ai-no-application, ml-criticism}. In fact, only a few medical areas, such as prosthesis~\cite{full-auto-pipline-presonalized-3d-print-hand} and dentistry~\cite{cad/cam}, tried to bind the process of 3D scanning, modeling, and printing. However, they all involved the use of CAD software which requires a certain level of expertise. Our goal is to not only build an automatic system that encompasses the full pipeline from scanning skin wounds to generating 3D-printable patches, but also provide a user-friendly interface that does not require special skills.

In addition, any ML system can fail especially while running in the wild; Bengio et al.~\cite{ai-iid} have reported that the performance of today's best ML systems often break down when they go out from the lab. For instance, some of the most popular and large image datasets posses geometric bias so that the ML models trained with those data resulted in errors with samples that are not included in the dataset~\cite{imagenet-geo-bias, ai-in-surgery-pros-perils}. Such errors can be disastrous in risk-sensitive fields like medical systems~\cite{ai-failure}. To minimize the dataset bias, we collected real world DFU data from multiple medical centers globally. 

Furthermore, the HITL ML approach has been proposed in multiple studies of medical image analysis~\cite{hitl-medical-survey} to maximize the performance while reducing the risk of malfunctioning. Amrehn et al.~\cite{uinet} utilized masks obtained from user scribbles to perform hepatic lesion segmentation and allowed users to add seed points if they were not satisfied with the prediction. Wang et al.~\cite{bifseg} proposed the image-specific fine-tuning method based on user scribbles. We adopt HITL to our user interface to reduce risks from unexpected predictions by allowing clinicians to modify and confirm the results. Also, our user-friendly interface and interaction do not demand separate learning process.

\section{Method}

\subsection{System Overview}
\begin{figure}[t]
  \centering
  \includegraphics[width=0.9\linewidth]{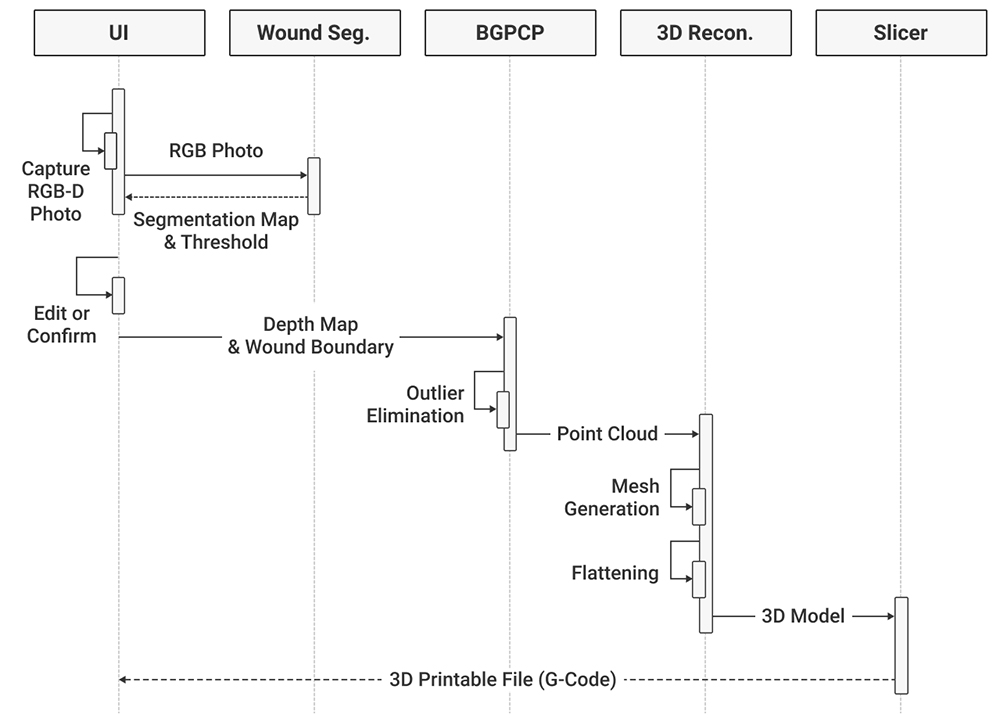}

   \caption{A sequence diagram of AiD Regen. Users can interact with the system via the user interface module and the rest of the modules seamlessly take care of the 3D-printable patch generation process.}
   \label{fig:archi}
\end{figure}

The \aid system encompasses the full-pipeline from capturing RGB-D images to generating 3D bio-printable patches to treat patients with DFUs. \aid only asks users to take a single photo and confirm the suggested wound boundaries without requiring technical knowledge about 3D modeling, and the rest of the complex patch generation process is taken care of by the system. The entire pipeline of the system is depicted in \cref{fig:archi}. When a clinician takes a photo with an RGB-D camera, the wound segmentation module analyzes the image and returns the detected wound boundary along with the segmentation map. The clinician can refine the result using the HITL-incorporated user interface. Then, the depth map and the refined boundary information are sent to the server. To increase the robustness against sensor noises near rough edges of the wound~\cite{depth-noise-need-filtering}, the boundary-guided point cloud processing (BGPCP) module eliminates outliers along the given boundary. The 3D reconstruction module generates a 3D surface mesh with the processed point clouds and then flattens it out to reduce printing time and cost. The final 3D model is created by using the flattened mesh and user-given thickness and fed to the slicer module to be converted into printing instructions known as G-code. By sending this G-code to a specified 3D bio-printer, the clinician can print out regenerative patches and treat the patients with DFUs.

\subsection{Capturing 3D Wound Images}

While various types of 3D cameras exist, the structured light sensors are known to have the highest precision when taken from a short distance~\cite{3d-image-review}. As the TrueDepth camera on Apple devices uses the structured light sensors and is reported to have better depth accuracy compared to other LiDAR or stereo-vision based sensors~\cite{ipad-truedepth-accuracy, 3d-image-review}, \aid utilizes the 12-megapixel front-facing camera of iPad Pro 11 (second generation) for capturing RGB-D wound images. The depth map is processed with temporal interpolation to smooth noises and remove holes and the resulting image consists of $2316 \times 3088$ RGB image and $480 \times 640$ depth map. Since taking a photo of a patient's wound using the front-facing camera is difficult, we mounted a silver-coated first surface mirror accessory onto the device (\cref{fig:teaser-a}).

\subsection{Wound Image Segmentation}
\begin{figure*}[t]
  \centering
  \includegraphics[width=0.85\linewidth]{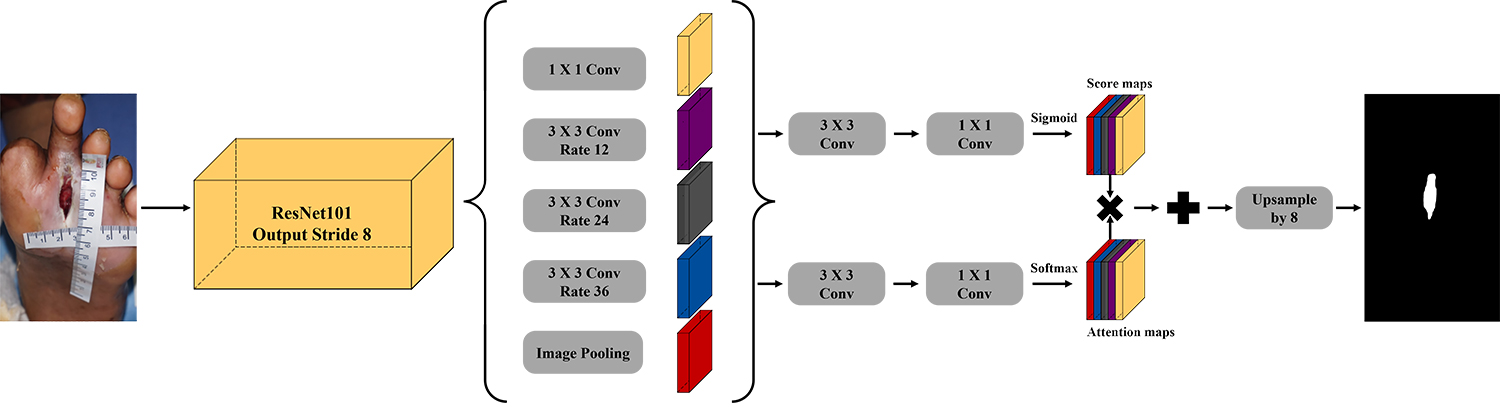}
  \caption{Our proposed model adopts the attention mechanism to \deeplab to preserve details of wound boundaries.} \label{fig:our_model}
\end{figure*}

Our first step towards training a deep-learning wound segmentation model was creating a DFU image dataset (\rokitdata). To reduce the dataset bias as much as possible, we collected data from institutional review boards (IRB) approved clinical trials conducted globally in four different medical centers: Foot and Ankle Specialists of Southwest Virginia (Virginia, USA), Korea University Guro Hospital (Seoul, Korea), Ankara City Hospital (Ankara, Turkey), and Hycare Super Specialty Hospital (Tamil Nadu, India). Photos were taken without any fixed conditions to reflect real-world uses. \rokitdata consists of 818 unique DFU images from 63 patients. The number of images per patient varies from 1 to 68 and the image size varies from $143 \times 176$ to $4128 \times 3096$. The images of the dataset not only contain DFU wounds but also other complications such as bleeding, slough, and necrosis. To prepare the dataset for the semantic segmentation task, two DFU experts independently labeled the data and then discussed until they reached a consensus. The dataset only contained foot focused images without any individual identifiers.

As we did not control the photo capturing process in order to reflect real-world scenarios, our dataset contains three different types of imbalances. First, DFU areas are sparse in images; only $3.4\%$ of the total pixels are classified as DFU and the rest as background. In this case, the majority background class dominates the net gradient, and thus, without careful data engineering, the imbalance can constrain the model from learning features of the minority DFU class which might be more important~\cite{class-imblance-backprop}. Second, the number of images per patient highly varies; about half of the images are from $11$ out of $67$ patients. Since feature extractors are likely to focus more on the wounds of the patient with greater number of images, generalizing a deep learning model might be difficult. Third, relatively small number of bleeding samples are included in the dataset. While DFU wounds can have many different visual characteristics like dark necrosis tissues, we found bleeding as one of the most critical complications of the wound since bleeding near the wound boundary can confuse the model from determining accurate boundaries. 

To handle the imbalances, we put several data augmentation stages as preprocessing. As the first stage, we resized input images in a way that the shorter edge of each image has $800$ pixels. Next, to balance the number of images per patient, we applied a deterministic oversampling method following \cref{eq:balance_aug_ref}:
\begin{equation} \label{eq:balance_aug_ref}
    S_{N_t}=\max(\min(C, N_M) - N_t, 0)
\end{equation}
where $N_M$ is the largest number of images that a single patient has in the dataset and $N_t$ is the current number of images of patient $t$. $C$ is a clamping parameter that sets the maximum number of oversampling images. $S_{N_t}$ is, finally, the number of images to oversample for patient $t$. We set the clamping parameter $C$ to $20$.

As the third stage, we performed another oversampling to balance bleeding and non-bleeding wound images. In this stage, we manually scanned through the dataset to identify bleeding samples and empirically determined the amount of oversampling; we analyzed the validation losses and chose the best value. For all the oversampled images during the second and third stages, we applied random rotation, shear transformation, and scaling; the random rotation was ranged between \qty{30}{\degree} and \qty{330}{\degree}, the shear transformation matrix $\bigl( \begin{smallmatrix}
1&s_x\\ s_y&1
\end{smallmatrix} \bigr )$ spanned between $-0.25$ and $0.25$, and the scale factor ranged from $0.5$ to $2$.

As the fourth stage, we applied eight randomized augmentation techniques: adding Gaussian noise, blurring, swapping color channels, luminance modification, flipping, rotating, scaling, and applying shear transformation. As a result, eight additional images were created per image. This stage was performed before the training to efficiently manage the dataset versions and reproductivity of experiments.

The last stage was designed to handle the sparseness of DFU areas in the images; only \qty{3.4}{\%} of the total pixels are classified as DFU. In this stage, we randomly sampled two wound pixels and six background pixels from each image and took $256 \times 256$ patches centered around those pixels, guaranteeing a quarter of the patches to be wound samples. Different from the previous stages, these patches were generated every iteration during the training.

We used a CNN-based model, \deeplab~\cite{deeplabv3}, and applied the attention mechanism to it. The data-efficient characteristics of CNN-based models have shown their advantages in wound segmentation tasks using small datasets~\cite{wound-seg1, wound-seg2, wound-seg3}. The architecture of our model is presented in \cref{fig:our_model}. From atrous spatial pyramid pooling layer, we could obtain features reflecting different fields of view. Each feature goes through two parallel and identical networks, which consist of $3 \times 3$ convolution layer and $1 \times 1$ convolution layer in series, to calculate score maps and attention maps using sigmoid and Softmax activations, respectively. The resulting score maps represent the features of the corresponding fields of view, and the attention maps show which field of view is more important at each pixel. The final segmentation map is obtained by upsampling the attention-weighted sum of the score maps.

For training, we split our dataset into training ($646$ images), validation ($81$ images), and test ($91$ images) sets. We constrained images of a patient included in one set to not appear in the other sets. We pretrained our model using the subset of COCO 2017~\cite{coco-dataset} and VOC2012~\cite{voc2012-dataset}. Images in the training set was passed through our data preprocessing and the total of $15,255$ images were obtained. We set the batch size to $32$ patches from four different images and the model was trained using four V100 GPUs. We stopped the training when the validation loss did not decrease for $1,000$ iterations. The loss function was binary cross entropy and the stochastic gradient descent was used as an optimizer with learning rate of $6.8\times10^{-4}$ without scheduling. For regularization, drop rate was set to $0.4$, and L2 regularization coefficient was set to $2\times10^{-4}$. On test time, our model takes a 2D image as an input and returns the segmentation map and the default threshold value so that the users can review the result using our interactive user interface.

\subsection{Human in the Loop}
\begin{figure}[b]
    \centering
    \begin{subfigure}[ht]{0.3\linewidth}
        \includegraphics[width=\linewidth]{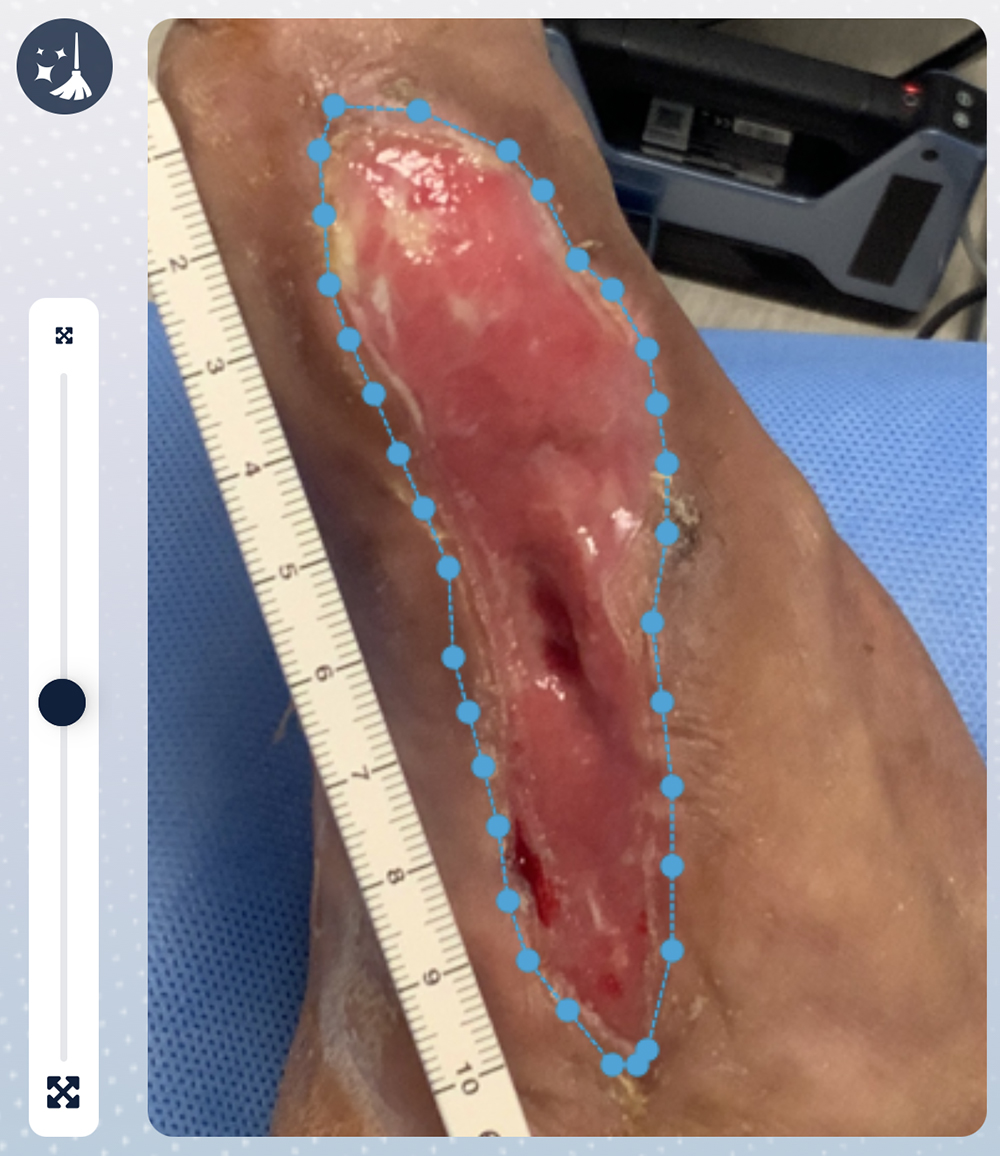}
        \centering
        \caption{}
        \label{subfig:ai_default}
    \end{subfigure}
    \begin{subfigure}[ht]{0.3\linewidth}
        \includegraphics[width=\linewidth]{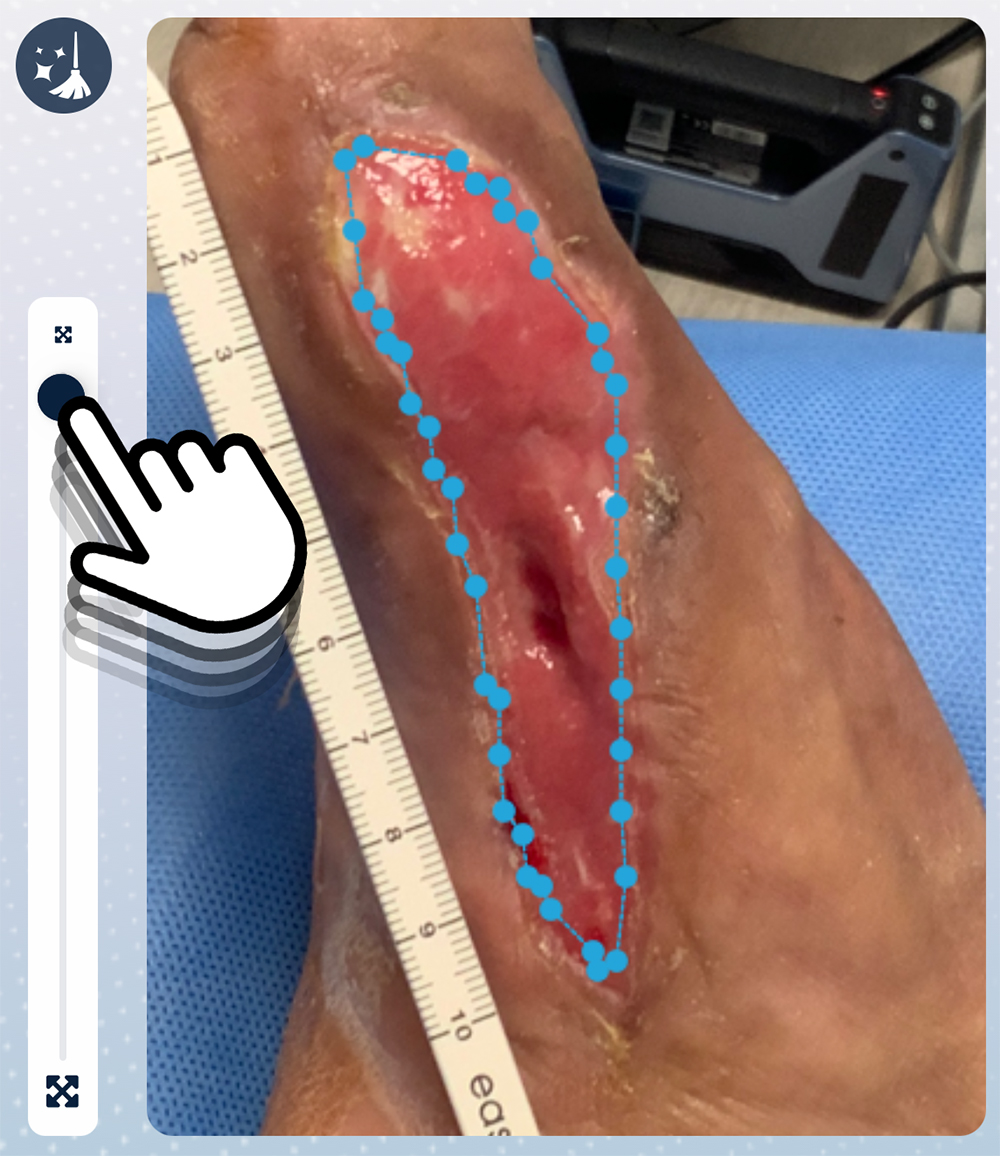}
        \centering
        \caption{}
        \label{subfig:ai_sen_up}
    \end{subfigure}
    \begin{subfigure}[ht]{0.3\linewidth}
        \includegraphics[width=\linewidth]{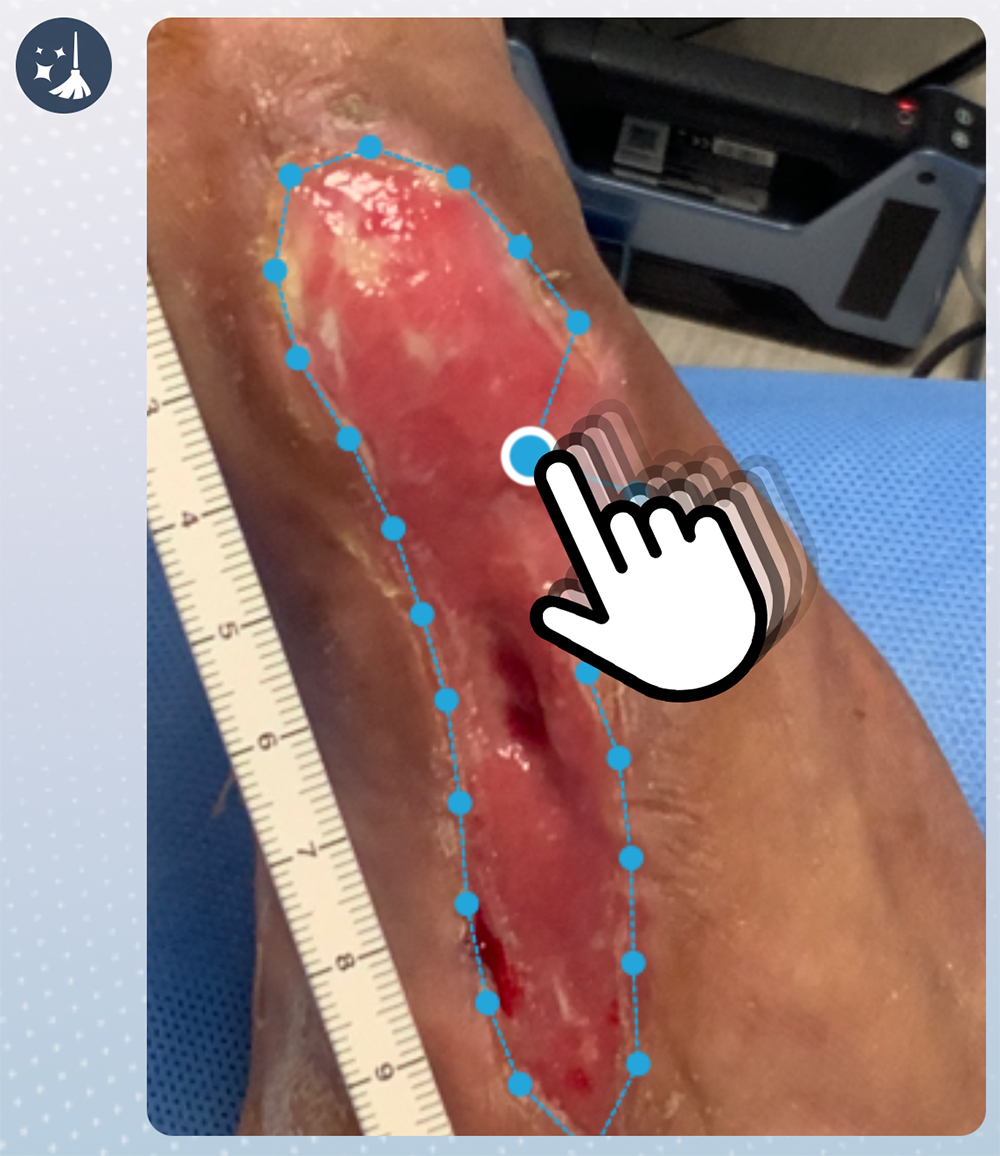}
        \centering
        \caption{}
        \label{subfig:drawing_modification}
    \end{subfigure}
    \caption{Our HITL-incorporated user interface. (a) \aid suggests a wound boundary at the user-provided seed point. (b) The boundary shrinks when the user moves the threshold slider upwards. (c) The user can finetune the boundary by moving each point or manually redraw the boundary in necessary.}
    \label{fig:our_ui}
    \raggedbottom
\end{figure}

No ML models are perfect and we have to admit that our model might occasionally produce unexpected predictions; however, such errors can be disastrous in the risk-sensitive medical field. Moreover, regardless of the accuracy of the prediction, clinical decisions can vary from clinician to clinician~\cite{interoperator-variability}. To maximize the performance of our model and handle inter-operator variability while preventing system malfunction, we adopted the HITL design to our user interface (\cref{fig:our_ui}). It consists of three major interactions: 
\begin{enumerate*}[label=\roman*)]
    \item seed point declaration,
    \item adaptive thresholding, and
    \item manual drawing or modification.
\end{enumerate*}

Firstly, when a user declares a seed point by touching any point within the wound area on the image, \aid displays a single wound area that includes the given seed point. Starting from a seed point is useful especially when multiple wounds exist in a single image and the clinician needs to confirm the intended target. When the deep learning model outputs a segmentation map, it usually applies fixed thresholds to decide the final classification. However, depending on the image, a different threshold value could give better result. \aid provides a slider so that the user can dynamically change the threshold while visually monitoring the changes in the segmentation result. Since the segmentation scores change gradually near the wound boundaries, when visualized, changing the threshold has the practical effect of shrinking or expanding the boundary (\cref{subfig:ai_sen_up}). 
While the user can simply accept the provided segmentation result without any additional interaction, \aid allows them to completely redraw the wound boundary or modify the given boundary by moving key points (\cref{subfig:drawing_modification}). Since all the drawing or modifications are done on a 2D image, our interactions are significantly easier to use compared to existing modeling solutions which require the use of complicated CAD tools.

\subsection{3D Model Generation}

\begin{figure}[ht]
    \centering
    \begin{subfigure}[ht]{0.3\linewidth}
        \includegraphics[width=\linewidth]{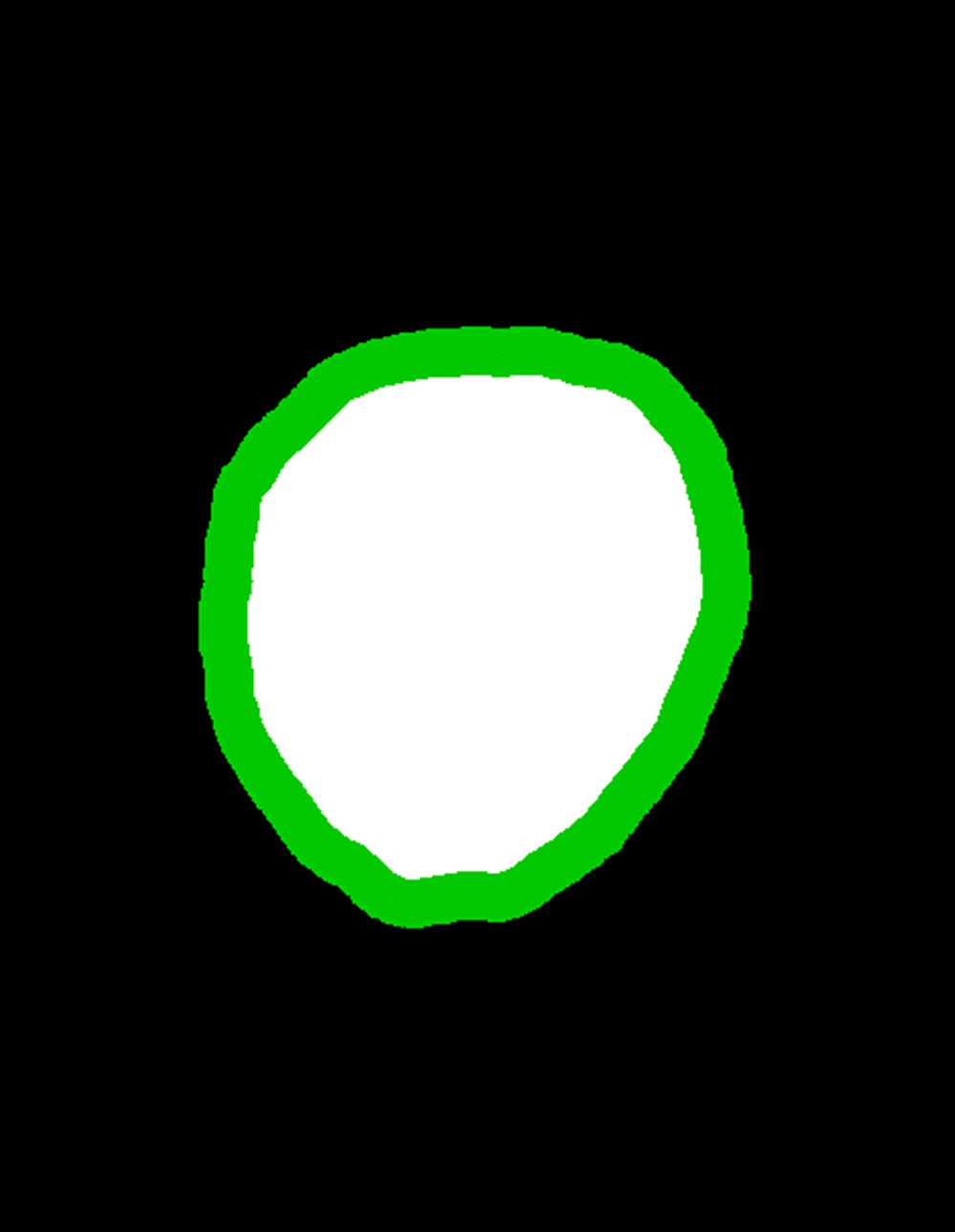}
        \centering
        \caption{}
        \label{subfig:bgpcp_seg}
    \end{subfigure}
    \begin{subfigure}[ht]{0.3\linewidth}
        \includegraphics[width=\linewidth]{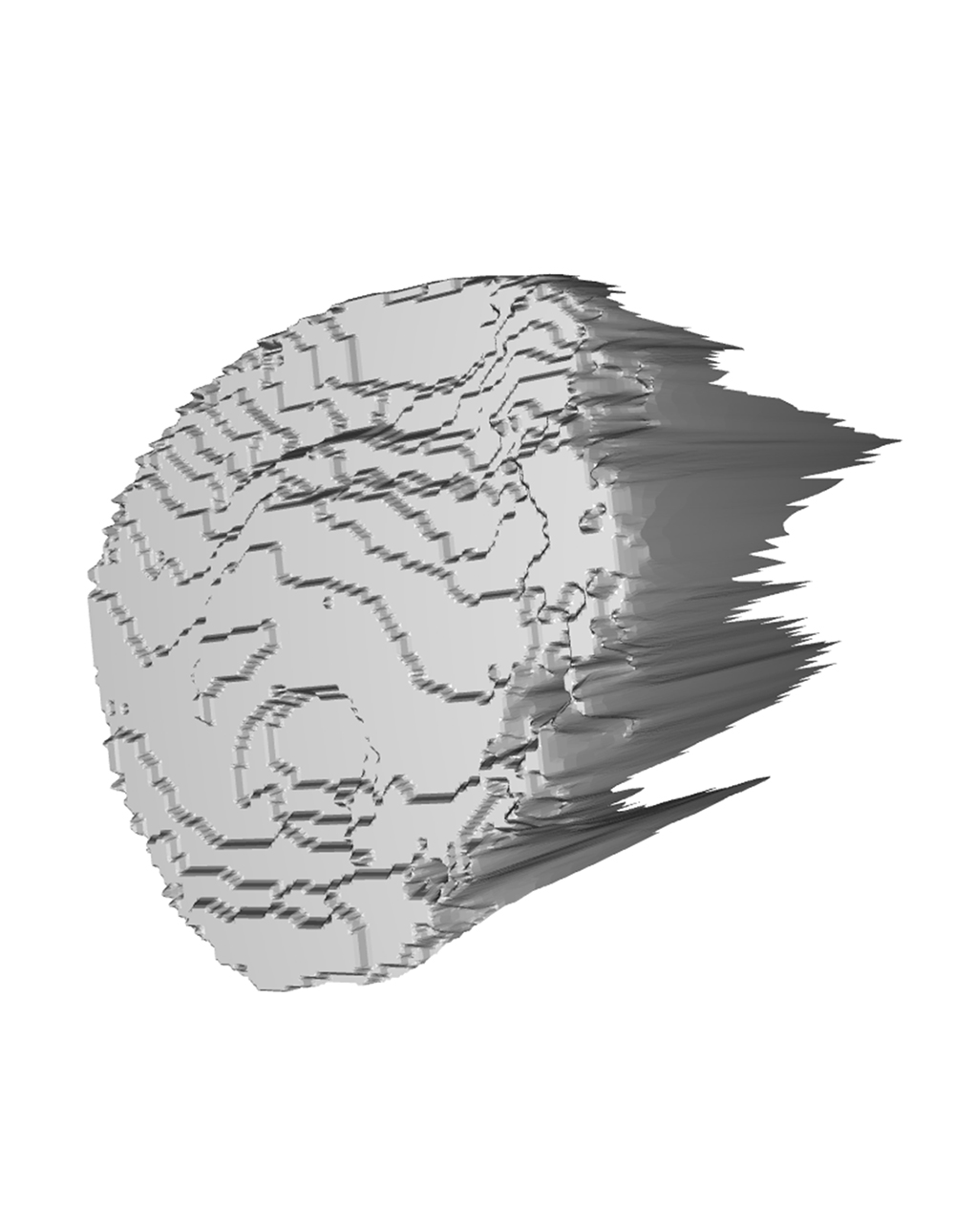}
        \centering
        \caption{}
        \label{subfig:bgpcp_ori}
    \end{subfigure}
    \begin{subfigure}[ht]{0.3\linewidth}
        \includegraphics[width=\linewidth]{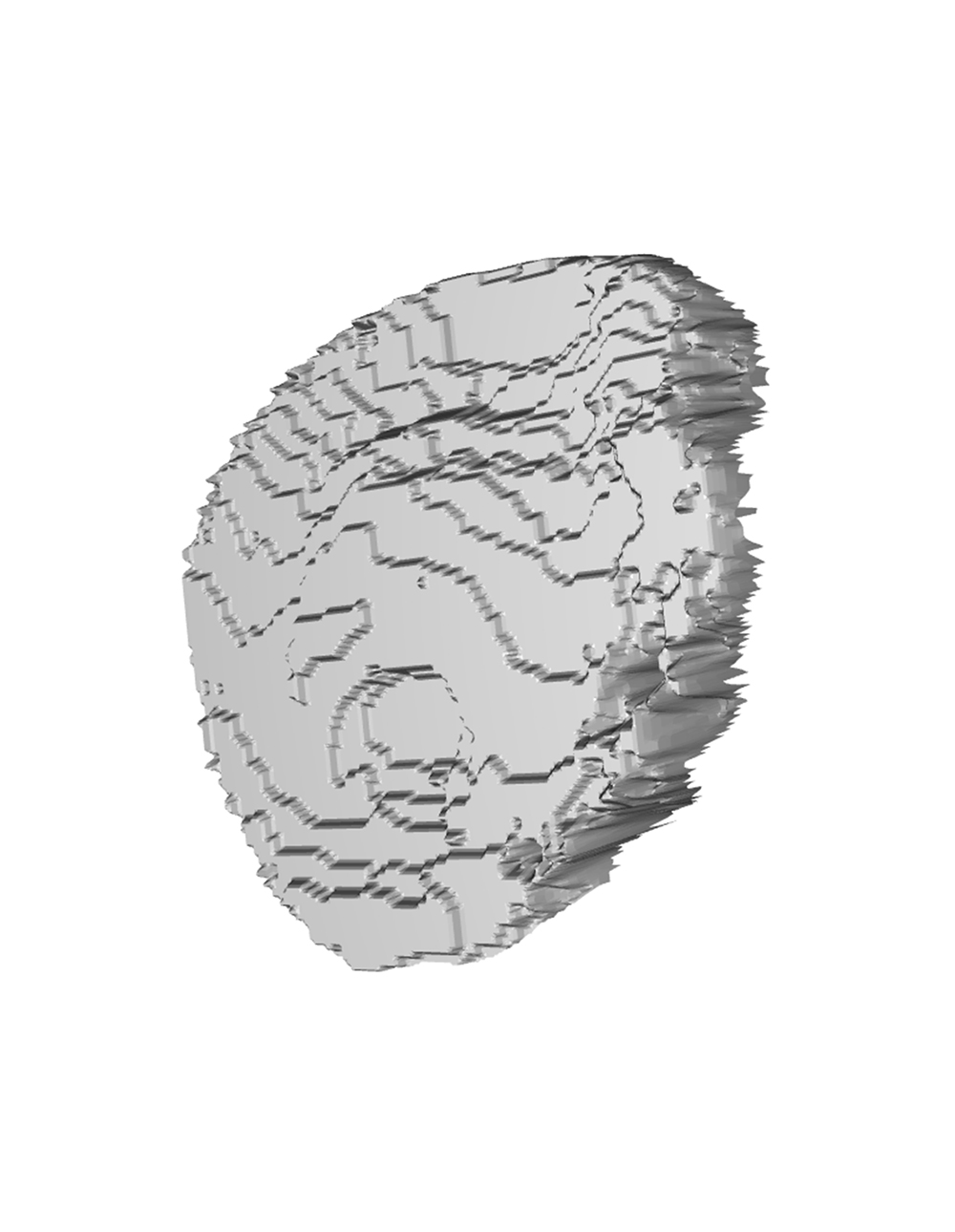}
        \centering
        \caption{}
        \label{subfig:bgpcp_processed}
    \end{subfigure}
    \caption{An example of BGPCP. (a) Green processing area is displayed on the segmentation map. (b) The original surface mesh with noise. (c) The BGPCP processed surface mesh.}
    \label{fig:bgpcp_result}
    \raggedbottom
\end{figure}

When the BGPCP module receives the wound boundary confirmed by the clinician, it takes the pixels that are within the given boundary to convert them into a point cloud. While we have obtained an accurate boundary via semantic segmentation and user interaction, depth noises near the edges are still present because the sensors can get confused when a point lies right in between the background and the wound~\cite{depth-noise-need-filtering}. Here, we suggest a BGPCP method that efficiently performs the radius based noise filtering~\cite{pcd-outlier-filtering} by limiting the processing area around the given wound boundary; we compute the distance transform~\cite{dist-transform} on the boundary and generate a thin processing area by setting the distance parameter to $0.1\times \sqrt{
A_{boundary}}$ where $A_{boundary}$ is the area within the boundary (\cref{fig:bgpcp_result}). The BGPCP method reduces the computation time of point cloud filtering by limiting the calculation area and is especially effective when the target wound is large.
 
Then, using the pixels that correspond to the points in the processed point cloud, we generate 2D mesh surfaces. We first create a convex-hull mesh using Delaunay triangulation and refine it to reflect the concavity of the wound area by removing all the edges whose midpoints lie outside of the boundary. The final 2D mesh is converted into a 3D mesh using the intrinsic matrix and depth map.

\subsection{Flattening and Slicing}
Flattening complex 3D meshes is critical in 3D bio-printing since taller models need higher supporters, consuming greater amount of filament and bioink. Exploiting the flexible characteristics of 3D regenerative patches made with MA-ECM, we flatten the mesh surface using the boundary-first flattening (BFF) algorithm~\cite{bff-algorithm}. Lastly, using the flattened mesh surfaces, we create a 3D patch model with the user-given thickness and export it as an STL file.

Commercial 3D printers usually have their own slicer software that converts 3D models into a printable format. In order to automate the entire pipeline, we ported a slicer, NewCreatorK~\cite{rokit-newcreatork}, that is intended to be used with a 3D bio-printer, Dr. INVIVO~\cite{rokit-drinvivo4d2}, to a cloud server connected with our gateway so that it can take an STL model as an input and return a G-code file as the output. Once the generated G-code is downloaded to the local device running \aid, the user can send it over to the bio-printer to physically print out the MA-ECM patch.

\section{Experiments}

\subsection{Segmentation Performance Comparison}

\begin{table}[b]
\centering
\resizebox{0.95\linewidth}{!}{%
{\tiny
\begin{tabular}{lccc}
\hline
\multicolumn{1}{c}{Model} & Recall & Precision & F1-Score \\ \hline
VGG16~\cite{wound-seg3}                     & 0.7835 & 0.8391    & 0.8103   \\
SegNet~\cite{wound-seg3}                    & 0.8649 & 0.8649    & 0.8503   \\
U-Net~\cite{wound-seg3}                     & 0.9129 & 0.8904    & 0.9015   \\
Mask-RCNN~\cite{wound-seg3}                 & 0.8640 & \textbf{0.9430}    & 0.9020   \\
MobileNetV2~\cite{wound-seg3}               & 0.8976 & 0.9086    & 0.9030   \\
MobileNetV2+CCL~\cite{wound-seg3}           & 0.8997 & 0.9101    & 0.9047   \\
DeeplabV3~\cite{deeplabv3}                 & 0.9207       & 0.9191           & 0.9199         \\
Our model                 & \textbf{0.9287} & 0.9172    & \textbf{0.9229}   \\ \hline
\end{tabular}
}%
}
\caption{Performance comparisons on the public wound dataset~\cite{wound-seg3}. Our model outperforms the previous models in recall and F1-score.}
\label{tab:wound-val-result}
\end{table}

We evaluated our model on the public wound dataset~\cite{wound-seg3} and compared the results against the previous models; since there was no performance report using \deeplab on the same dataset, we also trained and evaluated \deeplab for comparison. We firstly resized the training set to $800 \times 800$. We only applied the fourth stage of our data preprocessing as an augmentation without any oversampling or patch sampling. In the test step, the resulting segmentation map was downsized to $224 \times 224$ to match the original size of the label. As shown in \cref{tab:wound-val-result}, our model outperforms the previous models on recall ($0.93$) and F1-score ($0.92$).

We also evaluated the performance of our model using DFU-specific \rokitdata dataset. We trained U-Net~\cite{unet}, MobilenetV2~\cite{wound-seg3}, \deeplab~\cite{deeplabv3}, and our model by equally applying our data preprocessing and compared the performances on the validation set. As shown in \cref{tab:seg-val-result}, our model outperformed all the other models on mIoU ($0.88$) and F1-Score ($0.94$).

\begin{table}[t] 
\centering
\resizebox{0.95\linewidth}{!}{%
{\small
    \begin{tabular}{lccccccc}
        \hline
\multicolumn{1}{c}{Model}                     & S2 & S3  & HITL                  & mIoU   & Recall          & Precision       & F1-Score        \\ \hline
U-Net                     & \checkmark  & \checkmark  &                       & 0.7770 & 0.8733          & 0.8757          & 0.8745          \\
MobilenetV2               & \checkmark  & \checkmark  &                       & 0.8494 & 0.9242          & 0.9131          & 0.9186          \\
DeepLabV3                 & \checkmark  & \checkmark  &                       & 0.8707 & 0.9342          & 0.9277          & 0.9309          \\ \hline
Our model                 &    &    &                       & 0.5944 & 0.6153          & 0.9460          & 0.7456          \\
Our model                 & \checkmark  &    &                       & 0.6867 & 0.7075          & \textbf{0.9590} & 0.8143          \\
Our model                 &    & \checkmark  &                       & 0.8753 & 0.9286          & 0.9384          & 0.9335          \\
Our model                 & \checkmark  & \checkmark  &                       & 0.8840 & 0.9387          & 0.9382          & 0.9385          \\ \hline
Our model                 & \checkmark  & \checkmark  & \checkmark                     & \textbf{0.9043} & \textbf{0.9517} & 0.9478          & \textbf{0.9497} \\ \hline
    \end{tabular}
    }%
}
\caption{Results of our data processing and HITL interface on the validation set. S2 denotes the stage 2 augmentation which balances the number of images per patient, S3 denotes the stage 3 augmentation which oversamples the bleeding images, and HITL denotes that the HITL was simulated.}
\label{tab:seg-val-result}
\end{table}

Furthermore, to evaluate the effectiveness of our data preprocessing, we tested our model using the same dataset but omitting one or both of the stages two and three. As shown in \cref{tab:seg-val-result}, the model performed the worst with mIoU of $0.59$ when both stages were omitted and the best with mIoU of $0.88$ when all the data preprocessing was applied. Moreover, to demonstrate the maximum possible performance using the HITL interface, we simulated the best use by setting the optimal segmentation threshold for each image individually. We found that by using our HITL interface, the performance can be raised up to $0.90$ mIoU.

\begin{table}[b]
\centering
\resizebox{0.95\linewidth}{!}{%
{\tiny
\begin{tabular}{ccccc}
\hline
HITL & mIoU            & Recall          & Precision & F1-Score        \\ \hline
     & 0.8246          & \textbf{0.9465}          & 0.8650    & 0.9039          \\ \hline
\checkmark    & \textbf{0.8795} & 0.9329 & \textbf{0.9388}    & \textbf{0.9359} \\ \hline
\end{tabular}
}%
}
\caption{The performances of our model on the test set when the HITL was simulated or not.}
\label{tab:seg-test-result}
\end{table}

Lastly, we report the performance of our model on the test set in \cref{tab:seg-test-result}. The HITL interface increased the overall performance of mIoU, recall, precision, and F1-score.

\subsection{3D Model Accuracy}
\label{sec:exp-3d-model}
\begin{figure}[ht]
    \begin{center}
        \includegraphics[width=0.8\linewidth]{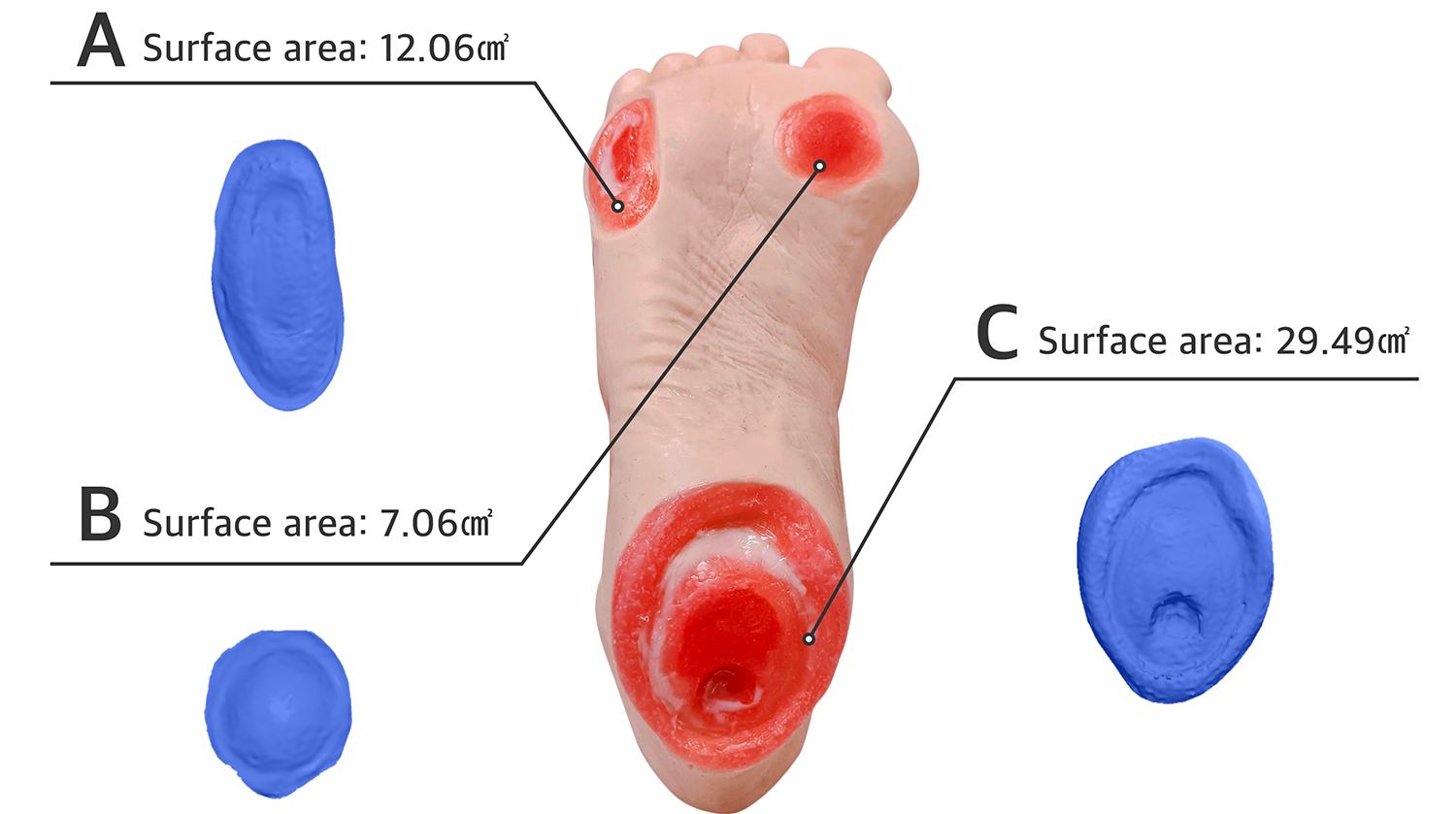}
        \caption{An artificial foot model with DFU wounds and the ground truth surface meshes.} \label{fig:dfu-model}
    \end{center}
\end{figure}

To measure the accuracy of 3D reconstructed wound model, we used an artificial foot model~\cite{fake-wound}, which has three different DFU wounds on the sole. As the ground truth, we scanned the model using an industrial 3D scanner, EinScan Pro 2X~\cite{einscan-pro-2x}, and measured the surface area of each wound (\cref{fig:dfu-model}). As an experiment, we captured RGB-D wound photos from a distance of \qty{20}{cm} at varying angles from \qty{-30}{\degree} to \qty{30}{\degree} in \qty{10}{\degree} increments. Five photos were taken from each angle and to focus solely on the reconstruction accuracy, we manually labeled each image to match the 3D scanned ground truth and compared the surface areas measured by \aid against the ground truth. The results show that \aid's reconstruction accuracy ranged from \qty{93.83}{\%} to \qty{97.07}{\%} per wound types and from \qty{94.75}{\%} to \qty{96.09}{\%} per angles (\cref{tab:3d-recon-acc}). The total mean absolute error was $\mu=0.92$~\unit{cm^2}, $\sigma=1.05$~\unit{cm^2} and the total average accuracy was \qty{95.27}{\%}.

\begin{table*}[t]
\tiny
\resizebox{\textwidth}{!}{%
\begin{tabular}{cccccc|ccccc}
\hline
     & \multicolumn{5}{c|}{MAE/Std (\unit{cm^2})}                    & \multicolumn{5}{c}{Accuracy (\%)}     \\ \hline
     
Type & 0°        & 10°       & 20°       & 30°       & Avr       & 0°    & 10°   & 20°   & 30°   & Avg   \\ \hline
A    & 0.55/0.34 & 0.37/0.41 & 0.88/0.87 & 0.52/0.45 & 0.59/0.61 & 95.40 & 96.91 & 92.72 & 95.69 & 95.15 \\
B & 0.19/0.1 & 0.2/0.08 & \textbf{0.17/0.07} & 0.26/0.13 & 0.21/0.11 & 97.29 & 97.17 & \textbf{97.66} & 96.29 & 97.07 \\
C    & 2.02/1.53 & 1.66/1.13 & 1.51/0.86 & 2.1/0.77  & 1.82/1.14 & 93.15 & 94.36 & 94.87 & 92.89 & 93.83 \\ \hline
Avg  & 1.2/1.38  & 0.77/0.97 & 0.85/0.9  & 0.96/0.96 & 0.92/1.05 & 94.75 & 96.09 & 95.08 & 94.95 & 95.27 \\ \hline
\end{tabular}%
}
\caption{The mean absolute error (MAE), standard deviation (Std), and accuracy of the wound surface reconstruction results are shown.}
\label{tab:3d-recon-acc}
\end{table*}

\subsection{Case Study}

\begin{figure}[ht]
\centering
\begin{subfigure}[b]{0.40\linewidth}
    \centering
  \includegraphics[width=\linewidth]{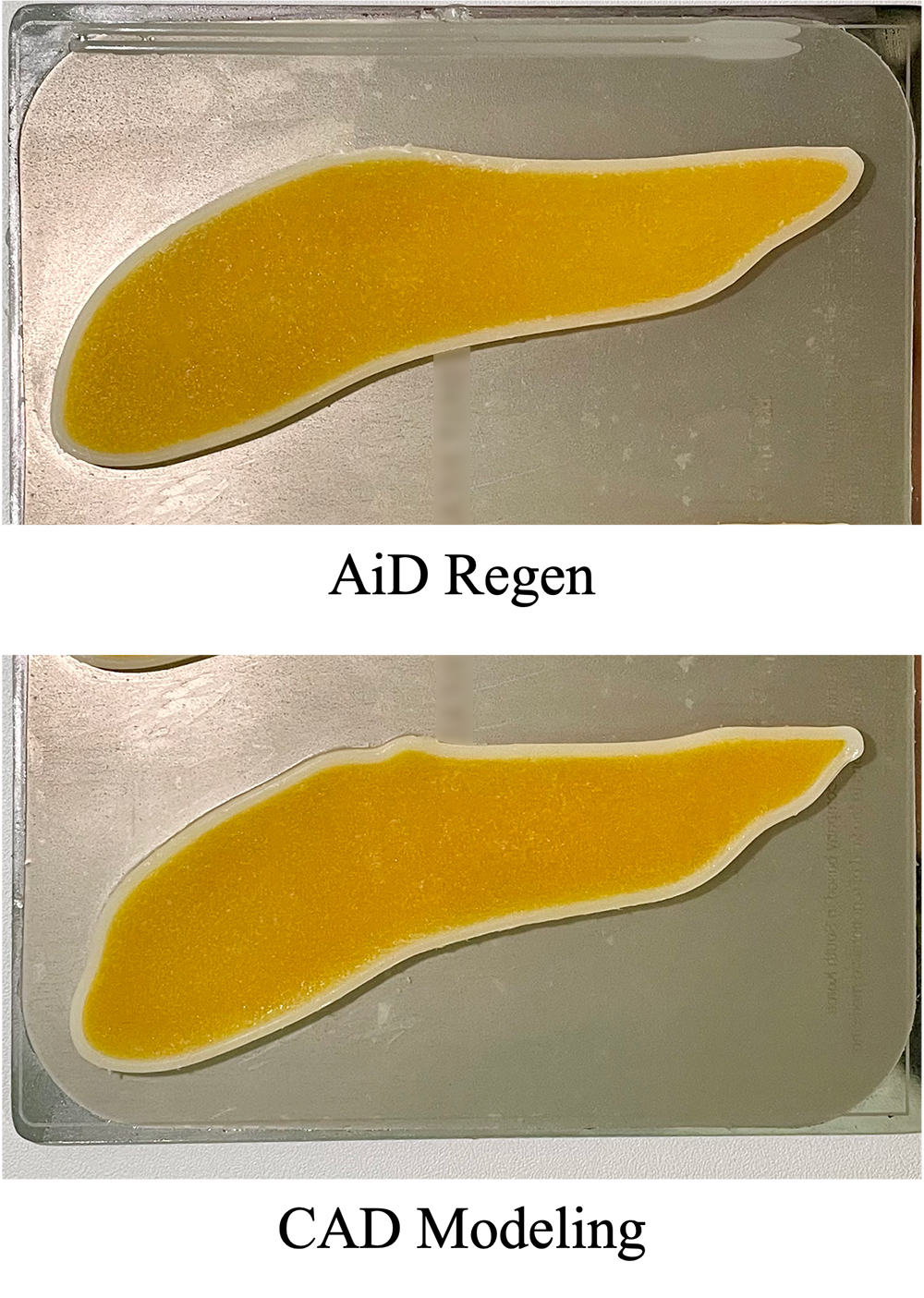}
  \caption{}
  \label{fig:patch} 
\end{subfigure}
\begin{subfigure}[b]{0.40\linewidth}
    \centering
  \includegraphics[width=\linewidth]{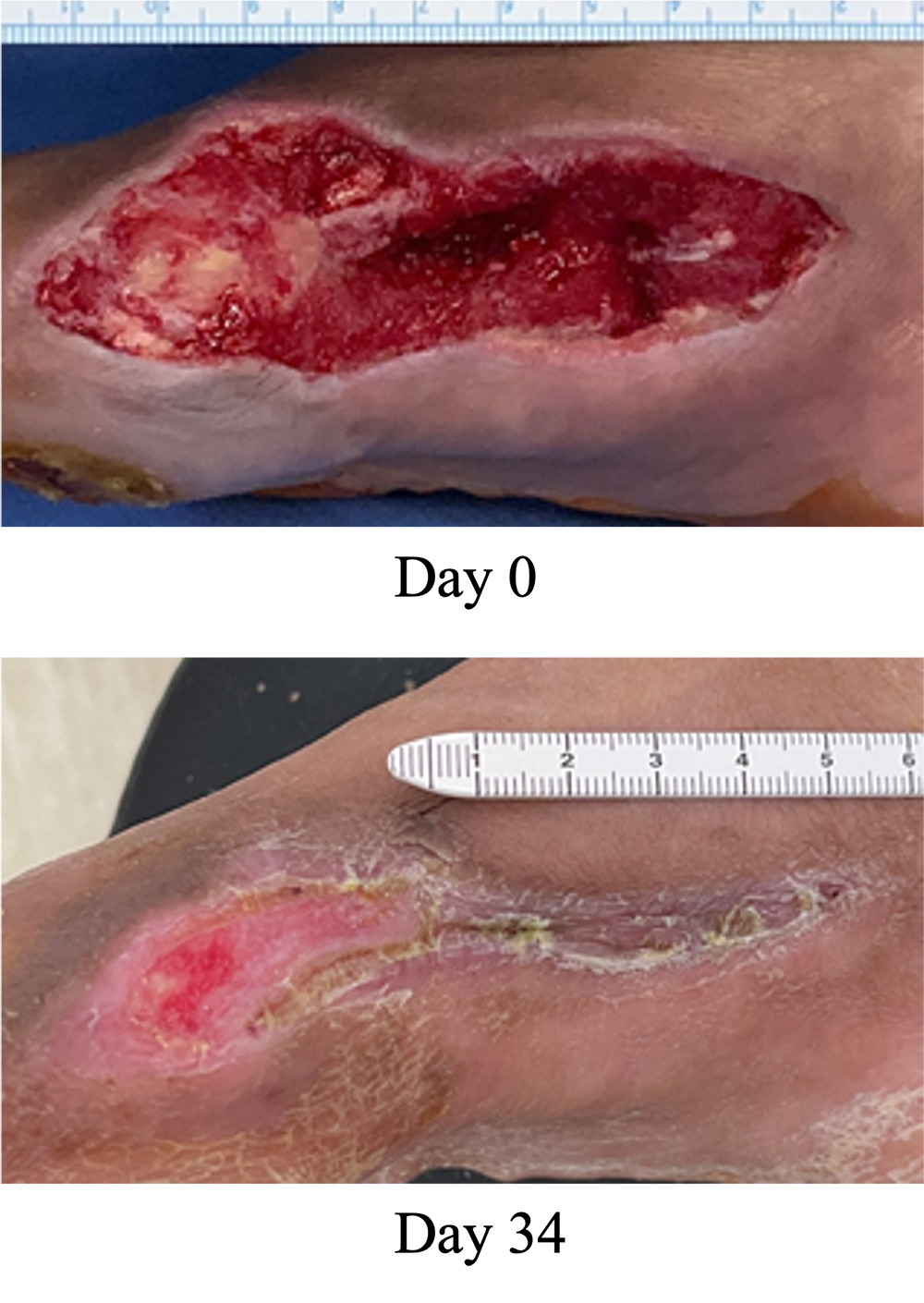}
  \caption{}
  \label{fig:wound}
\end{subfigure}

\caption{(a) The \aid generated patch (top) and the expert created patch using a CAD modeling tool (bottom) were very similar with only $2.03\%$ difference in volume, and (b) 96\% of the wound was healed within 34 days after the treatment.}
\label{fig:patch_and_wound}
\end{figure}

\begin{figure}[ht]
\centering
   \includegraphics[width=\linewidth]{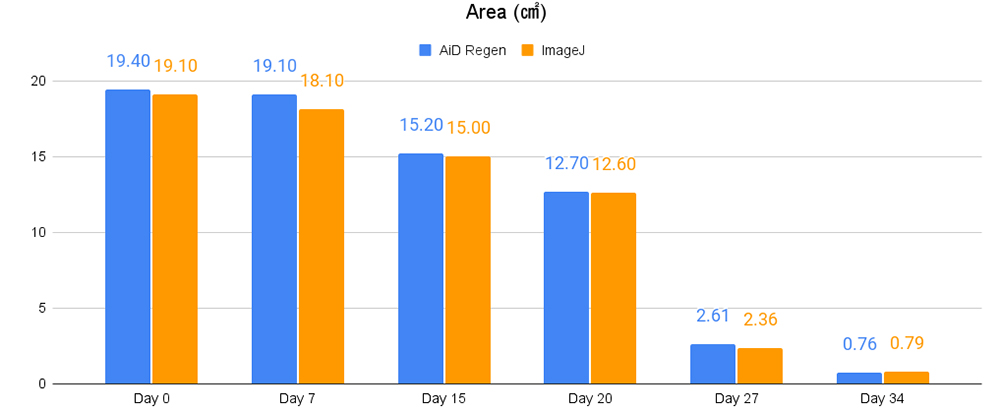}
\caption{The wound area measurements using \aid and ImageJ for every follow-up visit.}
\label{fig:imagej}
\end{figure}

We conducted a case study to demonstrate the effectiveness of \aid during a real clinical trial at the Catholic University of Korea Eunpyeong St. Mary's Hospital (Seoul, Korea. IRB PC21EOSE007). The patient was 67-year-old woman with Type 2 diabetes. A doctor generated a 3D patch model using \aid and printed an MA-ECM patch using a 3D bio-printer, Dr. INVIVO. As shown in \cref{fig:patch}, the patch generated by \aid was \qty{97.97}{\%} identical in area to the one created by an expert using a CAD modeling tool. The patient's wound was healed drastically as \qty{96}{\%} of the wound was closed within just 34 days after the treatment (\cref{fig:wound}). In addition, during each follow-up visit, we measured the wound size using both \aid and the conventional method which involves a tape measure and ImageJ~\cite{imageJ} software (\cref{fig:imagej}). The measurements of \aid and ImageJ were similar, both indicating the decreasing trend on every visit. \aid's measurements were on average \qty{0.30}{cm^2} greater that those of ImageJ.

\subsection{Discussion}

In order to build a robust ML model using small and imbalanced dataset, we introduced the multi-stage data preprocessing. The results in \cref{tab:seg-val-result} show that our data preprocessing successfully overcame such imbalances as our segmentation model achieved $0.8840$ mIoU when the full data preprocessing was applied; this is $29\%$ better than the one without applying stage 2 and 3 augmentations. In addition, our model, adopting the attention mechanism to \deeplab network in order to preserve details of the wound boundaries, outperformed previous models on both the public wound dataset~\cite{wound-seg3} as well as \rokitdata. 

Our HITL approach plays an important role in managing the segmentation performance as well as inter-operator variability; the boundary judgement and the treatment strategy can vary depending on the clinician. Our simple but highly customizable user interface allows clinicians to not only improve the segmentation results but also customize the results based on their judgement and preferences. The results in \cref{tab:seg-val-result} and \cref{tab:seg-test-result} indicate that the use of the HITL interface is capable of improving the overall performances. Furthermore, \aid only requires users to confirm the suggested boundary on the 2D image, removing the necessity of external training and reducing user's burdens compared to conventional CAD modeling. 

Unlike the conventional 3D scanning devices, \aid uses a single RGB-D image to reconstruct a 3D wound model by taking a hybrid approach that combines 2D segmentation and 3D reconstruction. While such hybrid approach using a single image significantly reduces computational complexities, the results in \cref{tab:3d-recon-acc} indicate that when DFU wounds are captured from \qty{20}{cm}-distance, \aid adequately reconstructed 3D models with average accuracy of \qty{95.27}{\%} and was robust against capturing angles of up to \qty{30}{\degree}. In the experiment \cref{sec:exp-3d-model}, the average surface area of the ground truth wounds was \qty{16.20}{cm^2} and the average MAE was \qty{0.92}{cm^2}. Assuming a circular shape, the \qty{95.27}{\%} average accuracy indicates the error range from \qty{-0.65}{mm} to \qty{0.63}{mm} in radius. Since the requirement we empirically received from the clinicians during the clinical trials was \qty{+-1.00}{mm} in radius, the performance of \aid is acceptable in practice.

\aid's hybrid approach connects the complex pipeline of multiple well-established techniques including deep learning based semantic segmentation, point cloud processing and filtering, triangular mesh generation, and mesh flattening. This carefully engineered system was able to achieve competitive performances and treat a real patient only with a small and unbalanced 2D image dataset. We hope our system and approach could encourage more researchers to apply established techniques and methods to solve practical real-world problems in various fields.

\subsection{Limitations}

While \aid's 3D reconstruction based on a single RGB-D image enhances user experience by simplifying the complex and time-consuming scanning procedures, \aid currently does not support curved wounds that cannot be captured in a single photo. In such cases, users inevitably have to take multiple photos from different angles and generate separate patches per angle. Thus, we are planning to develop a panorama-style capturing method that automatically determines the relationship between multiple images taken from different points-of-view. This way, users will be able to create a patch for complex wound surfaces in a single request.

Though \rokitdata was collected from various countries without any fixed condition to reflect the real-world environment, it is still not large enough and likely to be biased. That is, our hybrid approach based on the dataset showed impressive performance but might still produce unexpected results. To handle this limitation, we set up a continuous training that incrementally improves the model as the dataset grows with new wound images. In addition, while clinicians can determine 2D boundaries using our HITL interface, they currently cannot review and modify the final 3D model. Since we want to avoid recreating the burdensome experience of complex CAD-style modeling, we are attempting to develop an augmented reality interface so that users can intuitively manipulate 3D models in the real-world environment.

\section{Conclusion}

We introduced \aid, a novel system that generates 3D models based on the shape and geometry of wounds combining 2D image segmentation with 3D reconstruction so that they can be printed via 3D bio-printers during the surgery to treat DFUs. \aid seamlessly binds the full pipeline, which includes RGB-D image capturing, semantic segmentation, BGPCP, 3D model reconstruction, and 3D printable G-code generation, into a single system that can be used out of the box. We developed a multi-stage data preprocessing method to handle small and unbalanced DFU image dataset. \aid's HITL interface enabled clinicians to not only create 3D regenerative patches with a few touch interactions, but also customize and confirm wound boundaries. The results of the experiments indicate that our model outperformed prior wound segmentation models and our reconstruction algorithm was able to generate 3D wound models with compelling accuracy. Our case study on a real DFU patient demonstrated that \aid was effective in treating DFU wounds using a 3D bio-printer.
\section{Acknowledgement}
We thank Seokhwan You and Soon Yong Kwon for generous support and Gwanmo Park for practical discussion.
As participants of this study did not agree for their data to be shared publicly, supporting data is not available.

{\small
\bibliographystyle{ieee_fullname}
\bibliography{main}
}

\end{document}